\documentclass[article, onecolumn, 11pt, tightenlines, notitlepage, nofootinbib,longbibliography,floatfix, superscriptaddress]{revtex4-2}

\usepackage[utf8]{inputenc}
\usepackage{float}
\usepackage{graphicx}  
\usepackage{physics}
\usepackage{esint}
\usepackage{amssymb}  
\usepackage{mathtools}
\usepackage{amsmath}
\usepackage{amsthm}
\usepackage{mathrsfs}
\usepackage[colorlinks=true,linkcolor=blue,citecolor=blue,urlcolor=cyan,plainpages=false,pdfpagelabels]{hyperref}
\usepackage{color,xcolor,colortbl}
\usepackage{bm}
\usepackage[most]{tcolorbox}
\usepackage{enumitem}
\usepackage[normalem]{ulem}
\usepackage{diagbox}

\newcommand{\be}{\nopagebreak[3]\begin{equation}}
\newcommand{\ee}{\end{equation}}
\newcommand{\bfig}{\nopagebreak[3]\begin{figure}}
\newcommand{\efig}{\end{figure}}
\newcommand{\bea}{\nopagebreak[3]\begin{eqnarray}}
\newcommand{\eea}{\end{eqnarray}}

\numberwithin{footnote}{section}
\numberwithin{equation}{section}

\begin{document}

\title{Robustness of entanglement in Hawking radiation for optical systems immersed in thermal baths}

\author{Ivan Agullo}
\email{agullo@lsu.edu}
\affiliation{Department of Physics and Astronomy, Louisiana State University, Baton Rouge, LA 70803, U.S.A.
}

\author{Anthony J. Brady}\email{ajbrady4123@arizona.edu}
\affiliation{Department of Electrical and Computer Engineering, University of Arizona, Tucson, Arizona 85721, USA}

\author{Dimitrios Kranas}
\email{dkrana1@lsu.edu}
\affiliation{Department of Physics and Astronomy, Louisiana State University, Baton Rouge, LA 70803, U.S.A.
}

\begin{abstract}

Entanglement is the quantum signature of Hawking's particle pair-creation from causal horizons, for gravitational and analog systems alike. Ambient thermal fluctuations, ubiquitous in realistic situations,  strongly affects the entanglement generated in the Hawking process, completely extinguishing it when the ambient temperature is of the same order as the Hawking temperature. In this work, we show that optical analog systems have a built-in robustness to thermal fluctuations which are at rest in the laboratory. In such systems, horizons move relative to the laboratory frame at velocities close to the speed of light. We find that a subtle interplay between this relative velocity and dispersion protects the Hawking-generated entanglement---allowing ambient temperatures several orders of magnitude larger than the Hawking temperature without significantly affecting entanglement. 

\end{abstract}

\maketitle

\section{Introduction}
\label{intro}

That causal horizons spontaneously emit  pairs of entangled particles is a deeply fascinating prediction \cite{Hawking:1975vcx,unruh81} that subtly combines various aspects of physics: the kinematics of causal barriers, thermodynamics, and quantum mechanics. As a whole, the Hawking process (as we generically call this phenomenon, regardless of the physical system under consideration) is agreed to be a genuinely quantum process. However, which particular aspects of this process are quantum, and which are not? In a classical universe, there are two aspects of the Hawking effect which would not exist: (i) particle-creation starting from an initial vacuum (i.e., spontaneous emission) and (ii) quantum mechanically entangled Hawking pairs.

The first aspect is difficult to recreate, since it is challenging to completely isolate event horizons; thermal fluctuations and other sources of stochastic noise are ubiquitous. This is certainly true for astrophysical black holes, which are immersed in the cosmic microwave background radiation (amongst other source, such as the stochastic background of gravitational waves, the cosmic background of neutrinos, etc.). When embedded in a populated environment, a horizon emits induced or stimulated Hawking radiation, and it can be difficult to distinguish between spontaneous emission (originating from vacuum fluctuations) and induced emission.
 
What about the second aspect---entanglement? It turns our that background noise also deteriorates the Hawking-generated entanglement, even causing it to completely disappear for sufficiently intense noise. Mathematically, the core of Hawking's pair production is a process of {\em two-mode squeezing} (see, e.g., Refs.~\cite{Frolov:1998wf,jacobson2005qfcs}). In other words, the time evolution induces the following transformation, which relates creation and annihilation operators of the output modes to those of the input
\bea \hat a^{\rm(in)}_1&\to& \hat a^{\rm(out)}_1=\cosh r\,  \hat a^{\rm(in)}_1+e^{i\phi}\, \sinh r\,  \hat a^{\rm(in)\, \dagger}_2, \\ 
 \hat a^{\rm(in)}_2&\to& \hat a^{\rm(out)}_2=\cosh r\,  \hat a^{\rm(in)}_2+e^{i\phi}\, \sinh r\,  \hat a^{\rm(in)\, \dagger}_1. 
\eea 
where $r$ and $\phi$ are the squeezing intensity and squeezing angle, respectively. In this expression, $\hat a^{\rm(in)}_i$ ($i=1,2$) are the annihilation operators for the progenitors of Hawking pairs, while $a^{\rm(out)}_1$ labels the outgoing Hawking quanta and $a^{\rm(out)}_2$ their entangled partners. 

Under the above transformation, the vacuum evolves to a two-mode squeezed vacuum, in which case each subsystem is in a mixed thermal state, but there exist quantum correlations between the subsystems that purify the global state. If one replaces the initial vacuum by a thermal state of equal quanta $n_{\rm env}$ in each of the two input modes, it is easy to check that the entanglement in the final state degrades and disappears for $n_{\rm env}\geq e^r \sinh r$ (see, e.g., Appendix A of Ref.~\cite{brady2022prd} for details). Hence, when thermal fluctuations dominate, the squeezer is incapable of entangling the modes, and the final state is separable. This is a generic fact about two-mode squeezing processes---including the Hawking process of gravitational and analog systems.

The goal of this article it to provide a quantitative analysis for the impact that thermal noise has on Hawking-generated entanglement. For concreteness, we focus on an optical analog system containing a white-black hole pair that is moving at a finite velocity (a fraction of the speed of light) with respect to the lab frame. One of the main messages of our work is that this relative velocity introduces a large blue-shift, which in turn significantly increases the \emph{threshold temperature} (as measured in the lab) at which entanglement vanishes. The robustness of entanglement to high lab temperatures originates from (i) dispersive effects of the medium and (ii) the Lorentz boost between the lab frame---in which the equilibrium temperature of the thermal bath is naturally defined---and the comoving frame---in which the Hawking temperature is naturally defined. We argue that this is an interesting advantage for optical systems, absent in other analog systems for which horizons are at rest in the lab. We also argue that the robustness to thermal fluctuations has an interesting analogy to gravitational black holes.

Regarding previous works, the authors of Ref.~\cite{bruschi2013} identified the fragility of quantum entanglement to ambient noise and computed a threshold temperature for the analog Hawking effect in Bose-Einstein condensates (BECs). A similar analysis was performed in Refs.~\cite{agullo2022prl,brady2022prd} for optical analog systems assuming the rather unphysical situation of a thermal bath at rest with the propagating horizons. In either of these configurations, one does not find the enhanced robustness that we elucidate in this paper because there is no relative boost between the thermal bath and the horizons. That a relative boost can add robustness to some aspects of the Hawking process in optical analogs was pointed out qualitatively in Ref.~\cite{philbin08}, albeit with no reference to entanglement.

The rest of this article is organized as follows. Section \ref{optical} contains a brief summary of the Hawking effect in optical systems and includes references to more extended treatments where the interested reader can find additional details. Section \ref{lab} contains the main analysis of the paper---specifically, a quantitative study of the complex entanglement spectrum generated in the Hawking process and how such deteriorates in noisy environments. Section \ref{grav} contains a qualitative comparison with the physics of gravitational black holes. Conclusions and take-home points are gathered in Section \ref{conclusions}. Throughout, we adopt natural units in which $c=\hbar=1$.

\section{Optical analog white-black hole horizons}
\label{optical}

This section provides a  summary of the Hawking process in optical analog horizons \cite{philbin08}. The reader is referred to \cite{philbin08,linder16} for further details.

\begin{figure}[t]
    {\centering     
\includegraphics[width = 0.78\textwidth]{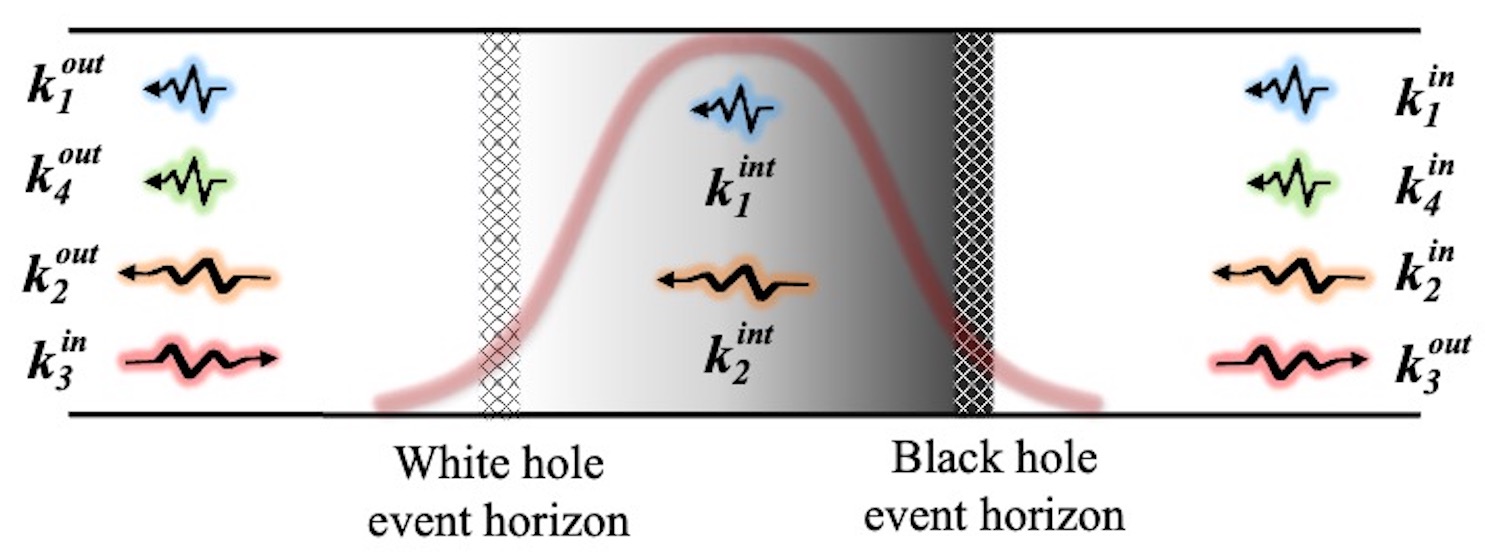}
}
\caption{Illustration of the structure of $in$, $int$, and $out$ modes for an optical analog white-black hole in the comoving frame (adapted from~\cite{agullo2022prl}). The analog white-black hole is generated by a strong electromagnetic pulse via the Kerr effect. There are four ingoing modes (three arriving at the black hole horizon and one at the white hole horizon), and four outgoing-modes. There are two real, propagating interior modes (int-modes) between horizons; the other two modes within this region are evanescent (i.e., they have complex wavenumber, and their amplitudes are exponentially suppressed).}\label{fig:modestructure}
\end{figure}

\subsection{The system}

Optical analogs~\cite{philbin08,drori19,rosenberg2020optical,Aguero-Santacruz:2020krw} rely on the Kerr effect, whereby a strong electromagnetic pulse propagating in an optical medium modifies the local refractive index. In turn, weak probe waves propagating thereon experience the perturbed refractive index near the pulse. Probes initially faster than the strong pulse will slow down when trying to overtake it and, for a strong enough pulse, its rear end becomes an impenetrable barrier~\cite{philbin08,petev2013blackbody,rubino2012soliton}. This is the analog of a white hole horizon. Similarly, an analog black hole horizon forms in the front end of the pulse. We thus have a pair of white- and black-hole event horizons that share an interior. We consider here that the material is a thin fiber and assume symmetry in the transverse $y$-$z$ plane. Restricting further to one fixed polarization, e.g. along the $y$ axis, the problem becomes effectively one dimensional~\cite{linder16}. 

From the perspective of an observer at rest in the laboratory, the pair white-black hole propagates with the strong pulse, which moves at group velocity $u$. The effective refractive index, therefore, depends on  space and time, $n_{\rm eff}=n+\delta n( x,t)$, where $n$ is the refractive index of the material in absence of any pulse, and $\delta n( x,t)$ is the perturbation generated by the strong pulse. It is more convenient to move to the frame comoving with the pulse because, in this frame, the pulse is static and the refractive index depends only on the spatial coordinate $\chi=\gamma(x-ut)$, where $\gamma=1/\sqrt{1-u^2}$ is the standard Lorentz factor. This, in turn, implies that frequencies of weak probes are conserved. We will denote  frequencies in the comoving frame by $\omega$ and the comoving wavenumber in the direction of propagation as $k(\omega)$. 
Due to the spatial dependence of the refrative index, the wavenumer $k$ is not conserved, and modes of same frequency $\omega$ but different wavenumbers $k_i(\omega)$ can mix or scatter; see Fig.~\ref{fig:modestructure} for an illustration.

\begin{figure}
    \centering
    \includegraphics[width=.5\linewidth]{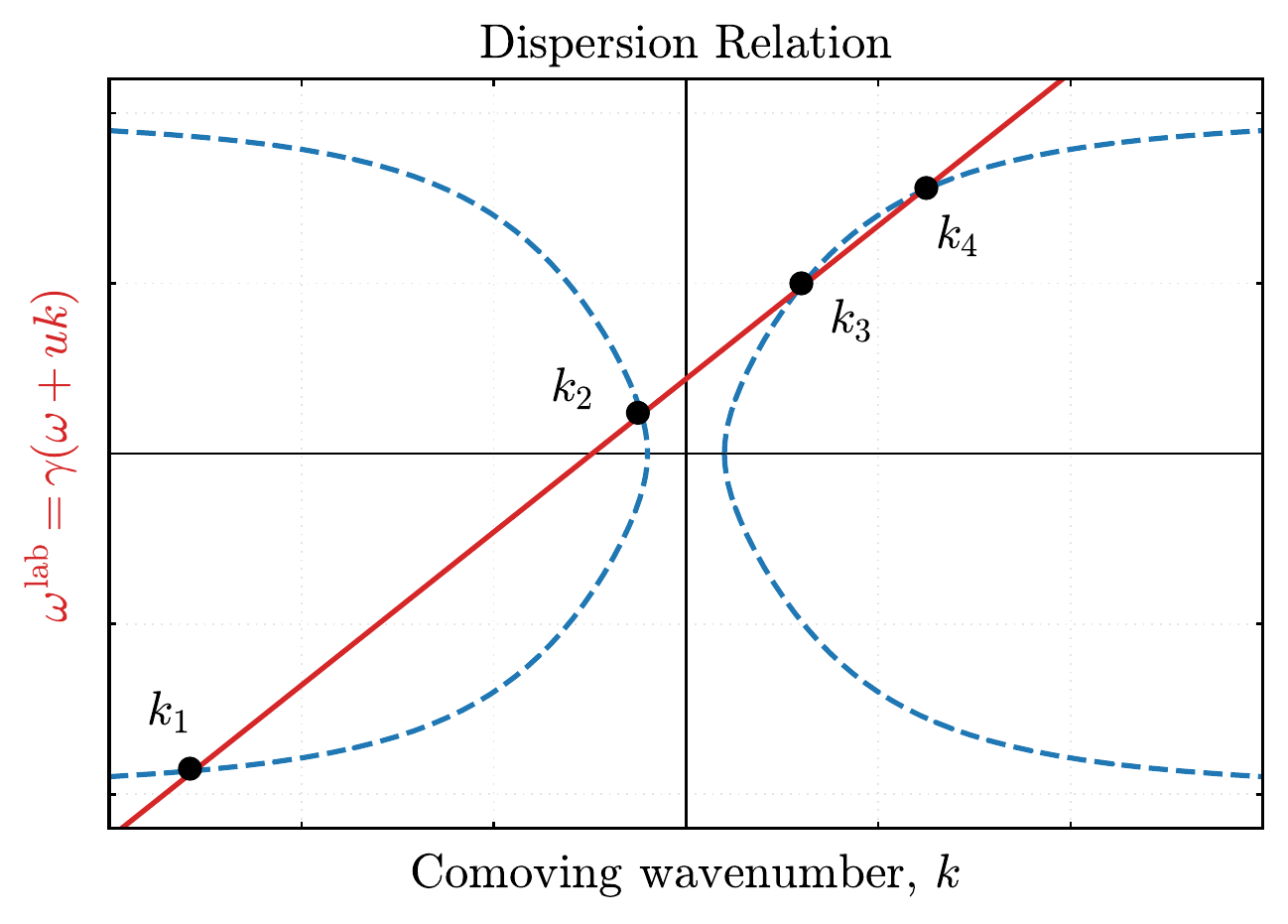}
    \caption{Graphical solution to the dispersion relation in the comoving frame for fixed comoving frequency $\omega$; see, e.g., Ref.~\cite{linder16} for explicit details. Allowed comoving wavenumbers $k_i$ are labelled. The central vertical and horizontal black lines indicate $k=0$ and $\omega^{\rm lab}=0$, respectively. The symplectic norm of a mode is determined by the sign of its lab frame frequency $\omega_{\rm lab}$ (shown in the vertical axis); thus, the $k_1$ mode is the only ``negative norm'' mode in this model.}
    \label{fig:dispersion}
\end{figure}

For concreteness, we consider the microscopic model proposed in Ref.~\cite{linder16} (building upon previous work~\cite{belgiorno15}), which describes a dielectric medium with one Sellmeier pole (i.e., with a single absorption resonance; see, e.g., Ref.~\cite{fox02} for an introduction to optical properties of dielectric media). Figure~\ref{fig:dispersion} illustrates a graphical resolution of the dispersion relation for a fixed comoving frequency $\omega$ far away from the strong pulse. There are four real solutions, which we  denote  by $k_i$, $i=1,2,3,4$. The modes $k_1$ and $k_4$ are short-wavelength modes, while $k_2$ and $k_3$ are of the same order as $\omega$.  The modes  $k_1$, $k_2$ and $k_4$ are left-movers in the comoving frame (negative group velocity, using the convention that the white hole is on the left of the black hole; see Fig.~\ref{fig:modestructure}). The mode $k_3$ moves to the right. One important property of the modes is the sign of their lab-frame frequency ${\rm sgn}({\omega}^{(\rm lab)})$, where ${\omega}^{(\rm lab)}=\gamma(\omega+ uk)$, because it determines the sign of the symplectic norm of the modes.\footnote{In general, for all complex solutions $f(\vec x,t)$ to the equations of motion, the symplectic norm is defined from the symplectic product via $(f_1,f_2)\equiv i\, \int_{\Sigma} d^3\Sigma\, n^{\mu}\,  (\bar f_1  \nabla_{\mu}f_2- \nabla_{\mu}  \bar f_1\,f_2)$, where $d^3\Sigma$ is the volume element of  the equal-time  hypersurface $\Sigma$, and $n^{\mu}$ its future-directed unit normal (see, for instance \cite{Wald:1995yp,fabbri05}).} The mode $k_1$ is the only mode with negative ${\omega}^{(\rm lab)}$, as it can be seen directly from Fig.~\ref{fig:dispersion}. This is crucial to understand the origin of spontaneous particle creation.

If the system is illuminated with a (ingoing) wave packet sharply centered at frequency $\omega$ and which has contribution from only one of the four possible wave-numbers $k_i$, the scattered (outgoing) wave will be a wave packet centered at the same frequency $\omega$ but will have contributions from all possible wavenumbers. When the dynamics mixes modes with the same sign of their symplectic norms, 
we have a simple scattering process; i.e., intensities of the incoming modes get distributed among the outgoing modes, but  the total intensity is conserved. However, when modes with different signs of their symplectic norm mix (e.g., a positive norm mode and a negative norm mode), the process classically corresponds to an {\em amplifier}, in the sense that the scattered waves generically are more intense than the incident ones (with energy supplied by the strong pulse, similar to how the mass sources Hawking radiation in semi-classical black holes). Quantum mechanically, a scattering process between two modes is described by a beam splitter, while an amplification process corresponds to a two-mode squeezer. Two-mode squeezing creates entangled quanta. Thus, for the optical analogs, any scattering process involving the $k_1$ mode (the only mode with negative norm) will generally lead to creation of entangled pairs.

\subsection{Formalism}

In a previous work \cite{agullo2022prl}, we have solved the dynamics of the system numerically---again, using the model of Ref.~\cite{linder16}---and derived the evolution matrix describing the dynamics (see Refs.~\cite{rubino2012soliton,petev2013blackbody,Gaona-Reyes:2017mks,Moreno-Ruiz:2019lgn,macher2009,Busch:2013gna,Finazzi:2010yq,Finazzi:2011jd,Finazzi:2012iu,Michel:2014zsa,Michel:2016tog} for numerical efforts in similar systems). The evolution takes a simple form in the Heisenberg picture, as we now describe. Since comoving frequency $\omega$ is conserved, it suffices to focus on one frequency at a time. For each frequency $\omega$, there are eight total asymptotic modes---four ingoing modes moving towards the horizons which scatter to the four outgoing modes moving away from the horizons; see Fig.~\ref{fig:modestructure}.

Let us denote by $\hat a_{k_i}^{\rm in}$ and $\hat a_{k_i}^{{\rm in}\,  \dagger}$ the annihilation and creation operators for wave-packet modes peaked at each of the four ingoing modes. From them, we can construct canonical pairs, $\hat Q_{k_i}\equiv \frac{1}{\sqrt{2}}\, (a_{k_i}^{\rm in}+\hat a_{k_i}^{{\rm in}\,  \dagger})$ and $\hat P_{k_i}\equiv -\frac{i}{\sqrt{2}}\, (a_{k_i}^{\rm in}-\hat a_{k_i}^{{\rm in}\,  \dagger})$. We collect these operators in a (column) vector
\begin{equation}\label{eq:R_in}
    \hat{\bm{R}}^{\rm(in)}\equiv\\\left(\hat{Q}_{k_1}^{\rm(in)},\hat{P}_{k_1}^{\rm(in)},\hat{Q}_{k_2}^{\rm(in)},\hat{P}_{k_2}^{\rm(in)},\hat{Q}_{k_3}^{\rm(in)},\hat{P}_{k_3}^{\rm(in)}, \hat{Q}_{k_4}^{\rm(in)},\hat{P}_{k_4}^{\rm(in)}\right)^\top,
\end{equation}
If we denote by  $\hat{\bm{R}}^{\rm(out)}$ the vector similarly constructed from the outgoing modes, Heisenberg evolution can be recasted as an expression for $\hat{\bm{R}}^{\rm(out)}$ in terms of $\hat{\bm{R}}^{\rm(in)}$. Although the existence of optical horizons originates in non-linear optics, the evolution of weak probes is well approximated by {\em linear} equations, with the non-linearities induced by the strong pulse all encoded in the optical properties of the medium, in analogy with the quantum field theory in curved spacetimes used in Hawking's original  derivation. This implies that the relation we are looking for can be written in matrix form as
\be  \label{eq:Rin_to_out}
\hat{\bm{R}}^{\rm(out)}= \bm{S}  \hat{\bm{R}}^{\rm(in)}\ee
where the $8\times 8$ matrix $\bm{S}$ contains all the information about the dynamics of the system. Furthermore, due to the linearity of the system, the matrix $\bm{S}$  can be obtained by solving the classical equations of motion. In other words, $\bm S$ contains the Bogoliubov coefficients between in and out complex solutions to the equations of motion.

In what follows, we shall focus on a special class of states, Gaussian quantum states (e.g., ground states of quadratic Hamiltonians, coherent states etc.), as these are easy to generate and manipulate in the lab. Moreover, it can be shown that linear dynamics [i.e., Eq.~\eqref{eq:Rin_to_out}] maps Gaussian states to Gaussian states. We use many efficient techniques for dealing with Gaussian states and refer to the reader to our previous work~\cite{agullo2022prl,brady2022prd} and Refs.~\cite{weedbrook2012, serafini17QCV} for further details about these techniques.

Gaussian states are completely characterized by their first and second moments (similar to multivariate Gaussian probability distributions). The first moments are codified in the vector  $\bm{\mu}\equiv\ev*{\hat{\bm{R}}}$, while the second moments are conveniently encoded in the covariance matrix 
\begin{align}
    \bm{\sigma}^{ij}&\equiv\ev{\left\{\hat{\bm{R}}^i-\bm\mu^i,\hat{\bm{R}}^j-\bm\mu^j\right\}},\label{eq:cov_general}
\end{align}
where the expectation value is computed with respect to the quantum state under consideration (either pure or mixed), and $\{\cdot,\cdot\}$ denotes the symmetric anti-commutator. For a Gaussian state, rather than working with a density operator $\hat \rho$, we can equivalently describe the state by a pair $(\bm{\mu},  \bm{\sigma})$. For linear systems, evolution preserves the Gaussian character of the state, and the relation between the in and out state is given by the matrix $\bm{S}$ discussed above

\begin{align} 
    \bm{\mu}^{\rm(out)}&=\bm{S}\cdot \bm{\mu}^{\rm(in)},\label{eq:out_mu_general}\\
    \bm{\sigma}^{\rm(out)}&=\bm{S}\cdot\bm{\sigma}^{\rm(in)}\cdot\bm{S}^\top.\label{eq:out_sigma_general}
\end{align}

Our primary focus in this work is on entanglement in optical analogs. To quantify such, we use the logarithmic negativity (LN) \cite{peres96,plenio05}. This is an easily-computable measure of entanglement for pure states and mixed states. For Gaussian quantum states (and other types of quantum states), the value of the LN has an operational meaning as the exact cost\footnote{The cost refers to exhange of a `currency'---which is entangled bits, or ebits---where 1 ebit is equal to the entanglement contained in 1 Bell pair.} that is required to prepare or simulate the quantum state under consideration \cite{wilde2020ent_cost, wilde2020alpha_ln}. Importantly, when restricting to Gaussian states and when one of the two subsystems is made of a single mode, the LN is greater than zero {\em if and only if} the state is entangled---regardless of the size of the other subsystem---and is larger for states with more entanglement. In other words, the LN is a faithful quantifier of entanglement for 1 versus $N$ mode bipartitions~\cite{weedbrook2012,serafini17QCV}. (We later apply this technical fact to optical event horizons to find threshold lab temperatures at which entanglement between subsystems strictly vanishes.)

Given a system $A$ of $N$ modes and system $B$ of $M$ modes occupying a Gaussian quantum state $\rho_{AB}$ with covariance matrix $\bm\sigma_{AB}$, the LN can be computed from the symplectic eigenvalues $\tilde \nu_I$ of the {\em partially transposed covariance matrix} ${\widetilde{\bm\sigma}_{AB}\equiv(\bm I_A\oplus\bm T_B)\bm\sigma_{AB}(\bm I_A\oplus\bm T_B)}$, where $\bm T_B=\bigoplus_{k=1}^M \big(\begin{smallmatrix}
  1 & 0\\
  0 & -1
\end{smallmatrix}\big)$ (see \cite{weedbrook2012,serafini17QCV} for further details),  
\begin{equation}
    {\rm LN}(\rho_{AB})=\sum_{J=1}^{M+N}\max\left[0, -\log_2(\tilde{\nu}_J)\right].
\end{equation}

\subsection{Analytical description of dynamics}

\begin{figure}
    \centering
    \includegraphics[width=0.7\textwidth]{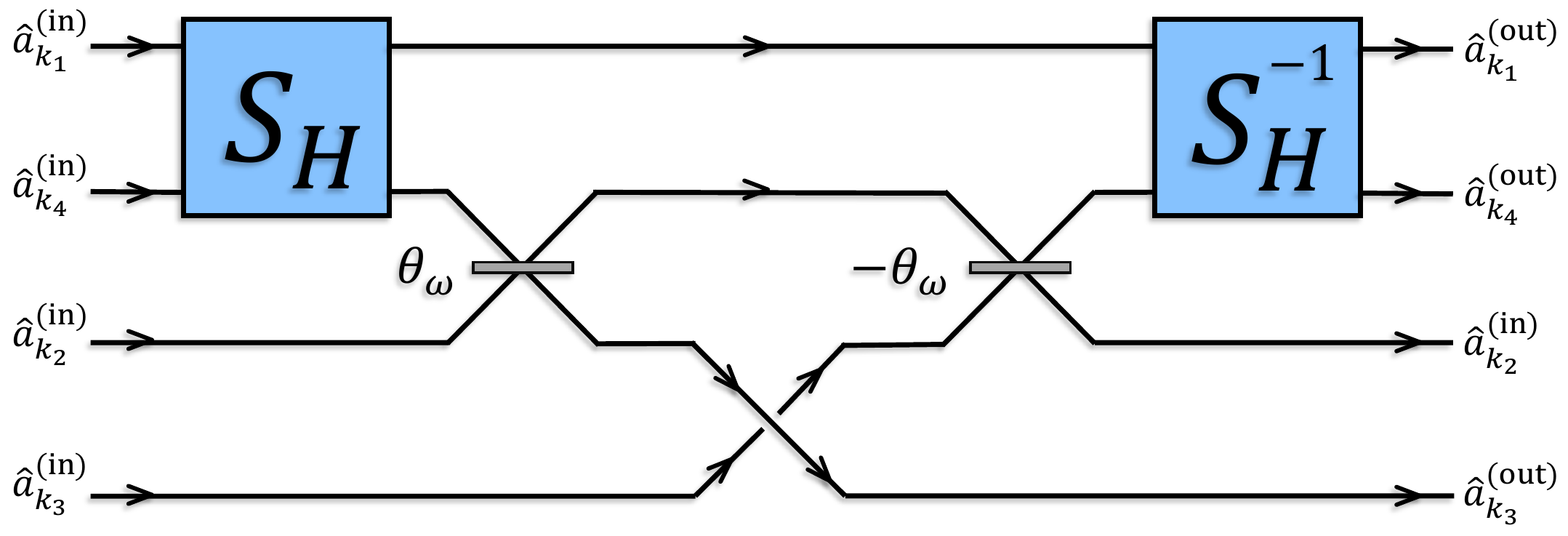}
    \caption{Symplectic-circuit of the Hawking process for an optical white-black hole pair (adapted from~\cite{agullo2022prl}). The  two-mode squeezer (blue box) and beam splitter (gray rectangle) on the left correspond to the black hole, while the elements to the right correspond to the white hole horizon.}
    \label{fig:wbh_circuit}
\end{figure}

To finish this summary, we consider a convenient way of describing the dynamics of the optical system under consideration. In previous works~\cite{agullo2022prl,brady2022prd}, we showed that the scattering process of the four ingoing modes to the four outgoing modes can be accurately described by a specific combination of two-mode squeezers and beam splitters; see Fig.~\ref{fig:wbh_circuit}. Each squeezer is associated with Hawking particle production from either the black hole horizon or the white hole horizon, which originates from scattering of the modes $k_1$ and $k_4$ to the modes $k_1$ and $k_3$. Since $k_1$ has negative symplectic norm, this scattering process is described by a two-mode squeezer. Each beam splitter describes the scattering between the modes $k_2$ and $k_3$ (and results in greybody factors). In realistic situations, such mixing is extremely weak and can be neglected, but we maintain this scattering in our analysis for completeness. The mixing between modes $k_2$ and $k_1$ (and likewise for modes $k_2$ and $k_4$) is negligibly small, due to the large differences in their wavelengths (see, e.g., Table~\ref{table:omegas}). 

Taking all this into account, the evolution from the ingoing to the outgoing modes can be described by the circuit of Fig.~\ref{fig:wbh_circuit}---we call this a \textit{symplectic circuit} because it is constructed from simple symplectic transformations \cite{agullo2022prl,brady2022prd}. From the symplectic circuit, it is straightforward to write an analytical expression for the evolution matrix of the system, ${\bm S}$, in terms of the parameters describing the squeezers and beam splitters; see Appendix~\ref{app:swb} for details. Such parameters can be extracted from numerical simulations. The circuit in Fig.~\ref{fig:wbh_circuit} and the analytical expression for  $\bm S$ is extremely useful to understand many aspects of the physics of the system. The full scattering matrix describing the transformation from the four ingoing modes to the four outgoing modes is
\begin{multline}
 \bm S_{\rm WB}=\\
\begin{pmatrix}
 (1+\cos ^2\theta \sinh ^2r_H)\bm{I}_2 &  \cos\theta\sin\theta\sinh r_H\bm{\sigma}_z & -\cos\theta\sinh r_H\bm{\sigma}_z & \cos ^2\theta \cosh{r_H}\sinh{r_H}\bm{\sigma}_z\\
 -\cos\theta\sin\theta\sinh{r_H}\bm{\sigma}_z & \cos^2\theta\bm{I}_2 & \sin\theta\bm{I}_2 & -\cos\theta\sin\theta\cosh{r_H}\bm{I}_2 \\
\cos\theta\sinh{r_H}\bm{\sigma}_z  & \sin\theta\bm{I}_2  & 0 & \cos\theta\cosh{r_H}\bm{I}_2 \\
 -\cos^2\theta\cosh{r_H}\sinh{r_H}\bm{\sigma}_z & - \cos\theta\sin\theta\cosh{r_H}\bm{I}_2 & \cos\theta\cosh{r_H}\bm{I}_2  & (\sin^2\theta-\cos ^2\theta\sinh^2r_H)\bm{I}_2
\end{pmatrix} ,\label{eq:SWB}
\end{multline}
where the subscript WB refers to ``white-black" hole, $\bm{I}_2$ is the $2\times2$ identity matrix, and $\bm{\sigma}_z$ is the Pauli-Z matrix. The squeezing intensity $r_H$ is a function of frequency and, for physical setups where the analogy with the Hawking effect is on firm ground, $r_H(\omega)={\rm arctanh} \, e^{-\omega/(2T_H)}$, where $T_H$ is the Hawking temperature associated to the event horizons. The angle $\theta$ is also frequency-dependent and is related to the greybody factor $\Gamma$ (typically introduced to describe classical scattering in the Hawking effect of black holes) via $\Gamma=\cos^2\theta$.  The scattering matrix $\bm S_{\rm WB}$ assumes that the strong pulse generating the horizons is symmetric, which amounts to say that their Hawking temperatures and greybody factors are the same. (It is straightforward to drop this assumption though.)

\subsection{Physical setup}

In our numerical simulations, we model the perturbation of the refractive index by a sech profile of the form ${\delta n(x,t)=\delta n_0\,\sech^2\left(\frac{t-x/u}{D}\right)}$. This is a common and convenient choice~\cite{philbin08, rubino2012soliton,drori19}. In this expression, $u$ is the group velocity of the perturbation caused by the strong pulse, and $x$ and $t$ are space-time coordinates in the lab frame. This profile depends on two real positive numbers, $\delta n_{0}$ and $D$, which determine the amplitude and width of the perturbation, respectively. References~\cite{agullo2022prl,companion} have explored values of $\delta n_{0}\in[0.01,0.1]$ and $D\in[2,10]$ fs. Away from these ranges, one does not have a horizon for all frequencies or dispersive effects are large and jeopardize the analogy with the Hawking effect. For the calculations in this paper, we will choose the concrete values $\delta n_0=0.05$ and $D=6\, \rm{fs}$. Other choices within the ranges specified above do not change our conclusions. Our tools permit to extend the analysis beyond these ranges, but analyzing such lacks motivation since we would be away from the analogy with the Hawking effect.  

For $\delta n_0=0.05$ and $D=6\, \rm{fs}$, the velocity of the horizons is $u=0.41$, and we obtain a value of $T_H=0.00046$ PHz ($T_H=3.51$ K if we restore the values of $\hbar$ and Boltzman's constant $k_B$) for the Hawking temperature. We extract this temperature form numerical simulations \cite{agullo2022prl,companion}.\footnote{Reference~\cite{linder16} contains an analytical expression for $T_H$ in terms of the parameters of the microphysical model, which provides an excellent approximation.} Table~\ref{table:omegas} contains the values of $k_i$ and lab frequencies for three representative values of the comoving frequency $\omega$, namely $\omega/T_H=10^{-1},\, 1\, , 10$. Observe the large blue-shift that the comoving frequency $\omega$ suffers when translated to the laboratory frame. For the modes $k_1$, $k_3$ and $k_4$ (the three modes primarily involved in the Hawking process), we have $\omega^{\rm lab}_i\gg \omega$. As a consequence, one would need temperatures in the lab frame much larger than the Hawking temperature $T_H$ in order to significantly populate these modes with thermal quanta. 

Although it is tempting to associate the large blue-shift exclusively with the Lorentz boost, dispersive effects play a key role. Notice that, for $u=0.41$, we have $\gamma=1.1$, which is close to unity. Consequently, if the system had the vacuum dispersion relation $k=\pm \omega$, the expression $\omega^{\rm lab}_{k_i}=\gamma\, (\omega+  u \, k_i)$ would imply that $\omega^{\rm lab}_{k_i}$ is of the same order as the comoving frequency $\omega$. For the optical system that we are considering, we find instead that $k_1$, $k_3$ and $k_3$ are all much larger than $\omega$ (see Table~\ref{table:omegas}), due to the bending of the dispersion relation away from the straight line $k=\pm \omega$; see Fig.~\ref{fig:dispersion}. In turn, this causes $\omega^{\rm lab}_i$ to be much larger than $\omega$ for $i=1,3,4$. On the other hand, the mode $k_2$ lives in a region in which the dispersion is close to vacuum dispersion and, consequently, its lab frequency is similar to its comoving frequency. In other words, there is no blue-shift for the $k_2$ mode; but the contribution of this mode to the Hawking process is faint. The results that we obtain in the next section originate from the blue-shift of $\omega^{\rm lab}_i$ (compared to $\omega$) for $i=1,3,4$.

\begin{table}
\renewcommand{\arraystretch}{1.7}

\begin{tabular}{c | c c c c} 
\hline\hline 

$\omega$  & $(k_1,\omega^{\rm lab}_{k_1})$ & $(k_2,\omega^{\rm lab}_{k_2})$& $(k_3,\omega^{\rm lab}_{k_3})$ & $(k_4,\omega^{\rm lab}_{k_4})$\\ 
\hline   
\hspace{.25em}$10^{-1}$\hspace{.5em} & \hspace{.25em}$(-1.62,-.737)\times10^4$ & $(-1.41,.459)\times10^{-1}\,\,$ & $(1.15,.530)\times 10\,\,$ & $(1.62,.737)\times10^4$ \\   
\hspace{.25em} $1$\hspace{.5em}  & \hspace{.5em}$(-1.63,-.739)\times10^4$\hspace{.25em} & \hspace{.25em} $(-1.41,.459)$ \hspace{.25em} & \hspace{.25em} $(1.15,.530)\times 10^{2}$ \hspace{.25em} & \hspace{.25em} $(1.62,.735)\times10^4$ \hspace{.25em} \\
\hspace{.25em} $10$ \hspace{.5em} & \hspace{.25em} $(-1.68,-.761)\times10^4$ \hspace{.25em} & \hspace{.25em} $(-1.41,.459)\times10$ \hspace{.25em} & \hspace{.25em} $(1.15,.532)\times 10^{3}$ \hspace{.25em} & \hspace{.25em} $(1.56,.709)\times10^4$ \hspace{.25em} \\   
\hline\hline
\end{tabular}
\caption{Comoving wavenumbers $k_i$ and lab frequencies $\omega_{k_i}^{(\rm lab)}$ for three representative values of the comoving frequency $\omega$. All quantities in units of the Hawking temperature $T_H=4.6\times10^{-4}$ PHz (in Kelvin, $T_H=3.51$ K). The numbers here are for a refractive index perturbation with $\delta n_0=0.05$ and $D=6\, \rm{fs}$, for which $u=0.41$.}\label{table:omegas}
\end{table}


\section{Thermal noise in the laboratory frame}
\label{lab}

We evaluate the impact that initial thermal fluctuations have on entanglement. We focus on thermal fluctuations since they are a common noise source in experimental settings. As explained above, we focus on situations in which the thermal bath is at rest in the lab frame and, thus, at rest with the optical medium. Due to the form of the dispersion relation, each mode $k_i$ ($i=1,2,3,4$) has a different population of thermal quanta; for a simplified analysis with isotropic noise, corresponding to a thermal bath at rest in the comoving frame, see Ref.~\cite{brady2022prd}.

\subsection{Ingoing and outgoing state}

In general, a thermal state for a quantum system of $N$ modes is a mixed state with density operator given by $\hat \rho=\, e^{-\beta\,  \hat H}/Z$, where $Z={\rm Tr} e^{-\beta \hat H}$ is a normalization factor (the partition function), $\hat H$ is the Hamiltonian of the system (which, here, is quadratic in the canonical variables), and $\beta=1/T$ the inverse temperature. This is, in fact, a Gaussian state and can be simply described by its first moments and its covariance matrix. The first moments of a thermal state are zero, and its covariance matrix is $\bm \sigma_{\rm th}=\bigoplus_{i=1}^N \, (1+2\, n_{\rm env;i})\, \bm I_2$, where $n_{\rm env;i}$ is the mean number of quanta for the mode $i$. For the optical system under consideration, the ingoing state of the modes is described by 
\begin{align}
    \bm \mu_{\rm th}^{\rm (in,\, lab)}&=\bm 0,\label{eq:therm_mu}\\
    \bm \sigma_{\rm th}^{(\rm in,\, lab)}&=\bigoplus_{i=1}^4 \, (1+2\, n^{(\rm lab)}_{\rm env;i}) \bm I_2\label{eq:therm_sigma}
\end{align}
in the laboratory frame, where $n^{(\rm lab)}_{\rm env;i}$ are given by the Bose-Einstein formula\footnote{We write $|\omega^{\rm lab}_{k_i}|$ rather than just $\omega^{\rm lab}_{k_i}$ because the mode $k_1$ has negative lab frequency for positive comoving frequency $\omega$. To compute the number of physical quanta, $|\omega^{\rm lab}_{k_i}|$ must be used in this expression.} 
\begin{equation}\label{eq:n_bose}
    n^{\rm (lab)}_{\rm env; i}=\frac{1}{e^{|\omega^{\rm lab}_{k_i}|/T_{\rm env}^{\rm (lab)}}-1}\, .
\end{equation}

The evolution problem of the last section was formulated in the comoving frame, for which frequencies $\omega$ are conserved. Therefore, we must translate the thermal state from the lab to the comoving frame. It is well known that the absence of correlations makes this an easy task; all we must do is transform  $n^{\rm (lab)}_{\rm env; i}$ to the comoving frame. In doing so, note that $n^{\rm (lab)}_{\rm env; i}$ is the number of quanta {\em per volume in phase space}. In other words, $n^{\rm (lab)}_{\rm env; i}\, {\rm d}^3x\, {\rm d}^3k$ is the number of quanta within ${\rm d}^3x$ and with wavenumber inside a volume ${\rm d}^3k$ centered at $\vec k$. Volumes in phase space are Lorentz invariant, which automatically implies that $n^{\rm (lab)}_{{\rm env};i}$ is a Lorentz scalar (see, e.g., \cite{Weinberg:2008zzc}). Hence, the number of quanta in the comoving frame in the mode $k_i$ is given by  $n^{\rm (co)}_{\rm env; i}=n^{\rm (lab)}_{\rm env; i}$. We can thus describe the quantum state in the comoving frame (via $\bm\mu^{\rm(in,\,co)}_{\rm th}$ and $\bm\sigma^{\rm(in,\,co)}_{\rm th}$) by simply relabeling $n_{\rm env;i}^{\rm (lab)}$ as $n_{\rm env;i}^{\rm (co)}$ in Eqs.~\eqref{eq:therm_mu} and \eqref{eq:therm_sigma}. 

Observe that the number of noisy quanta in the mode $k_i$, for a thermal bath at rest in the lab frame, is determined solely by the ratio $|\omega^{\rm (lab)}_i|/T^{\rm (lab)}_{\rm env}$. This is interesting because, although the modes $k_i$ have the same comoving frequency $\omega$, they have {\em very different} lab frequencies (see Table \ref{table:omegas}). This in turn implies that the modes are initially populated with noisy quanta in an exponentially different manner. In particular, the values of lab frequencies written in Table \ref{table:omegas} result in the hierarchy $n^{\rm (co)}_{\rm env; 1}\approx n^{\rm (co)}_{\rm env; 4} \ll n^{\rm (co)}_{\rm env; 3} \ll n^{\rm (co)}_{\rm env; 2}$. This hierarchy affects the entanglement between different subsystems, as we discuss in the next section.

 \begin{figure}[t]
    \centering
    \includegraphics[width=.6\linewidth]{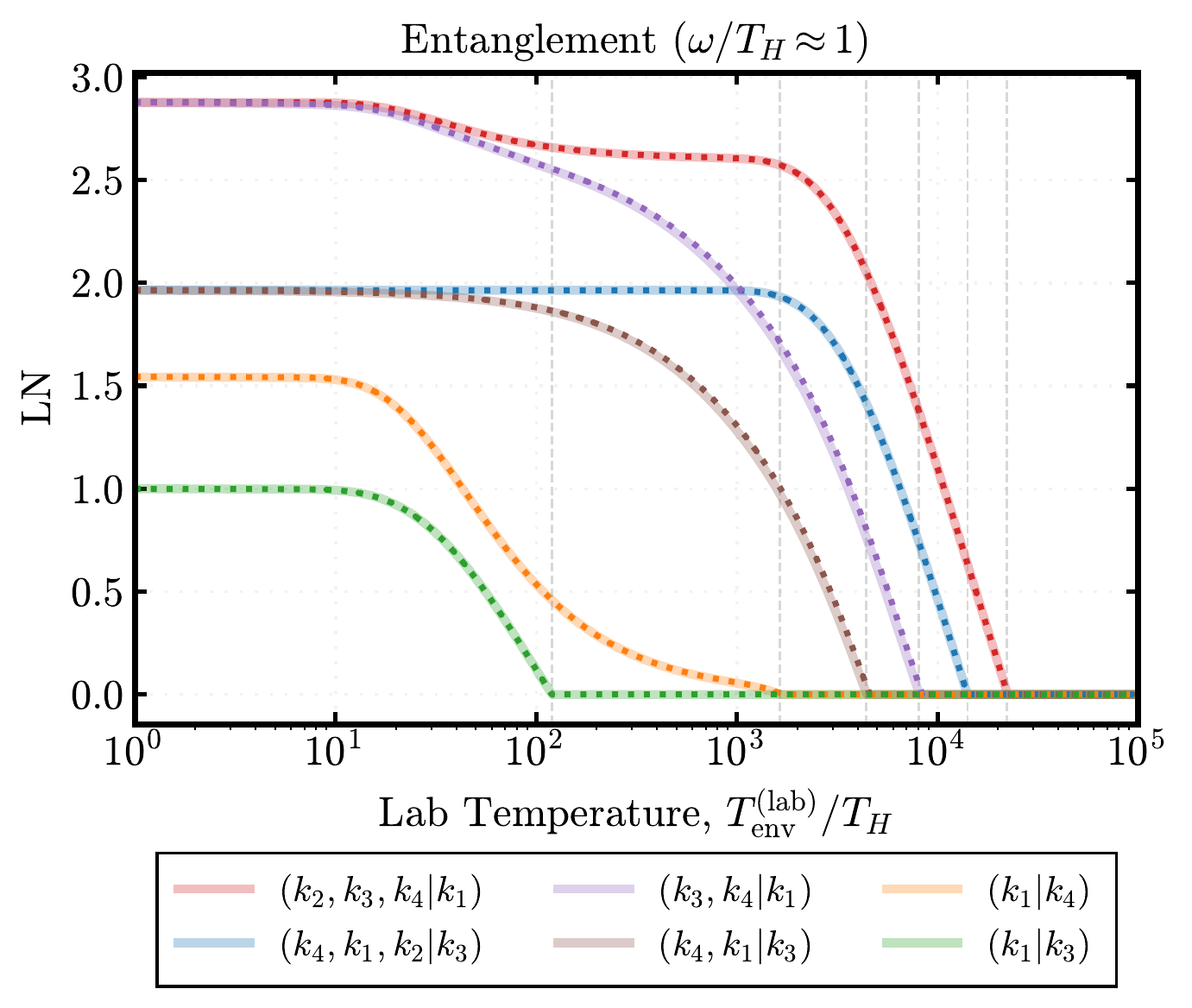}
    \caption{Entanglement versus lab temperature $T_{\rm env}^{(\rm lab)}$ for various bipartitions evaluated at $\omega/T_H\approx1$. Here $T_H=4.\times10^{-4}$ PHz (in Kelvin, $T_H=3.51$ K). Solid curves correspond to numerically obtained values. Dotted curves correspond to the analytic circuit approximation. The agreement is excellent. Vertical dashed lines label the threshold lab temperature $T_{\rm env}^{\star\,(\rm lab)}$ at which entanglement vanishes for that bipartition; see second row of Table~\ref{table:threshold}.}
    \label{fig:ln}
\end{figure}

Finally, given an initial thermal state in the lab frame with temperature $T_{\rm env}^{(\rm lab)}$ and using the general transformations~\eqref{eq:out_mu_general} and \eqref{eq:out_sigma_general}, the outgoing state, in the comoving frame, is characterized by
\begin{align} 
    \bm{\mu}^{\rm(out,\,co)}&=\bm{S}_{\rm WB}\cdot \bm{\mu}^{\rm(in\,co)}_{\rm th}=\bm 0,\label{eq:mu_out_wb}\\
    \bm{\sigma}^{\rm(out,\,co)}&=\bm{S}_{\rm WB}\cdot\bm{\sigma}^{\rm(in,\,co)}_{\rm th}\cdot\bm{S}_{\rm WB}^\top.\label{eq:sigma_out_wb}
    \end{align}
The covariance matrix $\bm \sigma^{(\rm out,\,co)}$ encodes information about all four outgoing modes; though, we often focus on particular subsystems by looking at the reduced moments---i.e., sub-matrices of $\bm \sigma^{(\rm out, co)}$. From the output covariance matrix, we can extract information about the number of particles created (see Appendix~\ref{app:particle_number} for analytical expressions for these quantities), the amount of entanglement generated in the scattering process (see below), etc. We remark that, although entanglement calculations in next section are done in the comoving frame, results do not change by going back to the lab frame because a Lorentz boost does not mix the modes. 

\subsection{Entanglement degradation}

\begin{table}
\renewcommand{\arraystretch}{1.6}
\centering

\begin{tabular}{c |c c c c c c c} 
\hline\hline 

\hspace{.75em}$\omega$\hspace{.75em} & \hspace{.75em}$(k_2, k_3, k_4|\,k_1)$ & \hspace{.75em}$(k_4, k_1, k_2|\,k_3)$ & \hspace{.75em}$(k_3, k_4|\,k_1)$  &  \hspace{.75em}$(k_4, k_1|\,k_3)$ & \hspace{.75em}$(k_1|\,k_2)$ &\hspace{.75em} $(k_1|\,k_3)$ &\hspace{.75em} $(k_1|\,k_4)$ \hspace{.75em}\\ 

\hline   
$10^{-1}$ & \hspace{.5em} $1.41\times10^6$  & \hspace{.5em} $1.40\times10^5$ &
\hspace{.5em} $9.51\times10^{3}$ &
\hspace{.5em} $620$ &
\hspace{.5em} $3.86\times10^{-3}$ & \hspace{.5em} $2.30$ & \hspace{.5em} $1.74\times10^3$ \\   
1 & \hspace{.5em} $2.22\times10^4$ &  \hspace{.5em} $1.41\times10^4$ &
\hspace{.5em} $8.06\times10^{3}$ &
\hspace{.5em} $4.42\times10^{3}$ &
\hspace{.5em} $4.42\times10^{-2}$ & \hspace{.5em} $1.20\times10^2$ & \hspace{.5em} $1.64\times10^3$\\   
10 & \hspace{.5em} $1.50\times10^3$ & \hspace{.5em} $1.50\times10^3$ &
\hspace{.5em} $1.41\times10^3$ &
\hspace{.5em} $1.41\times10^3$ &
\hspace{.5em} $1.27$ & \hspace{.5em} $1.40\times10^3$ & \hspace{.5em} $340$\\   
\hline\hline 
\end{tabular}
\caption{Threshold lab temperatures $T_{\rm env}^{\star\,(\rm lab)}$ at which entanglement for various bipartitions vanishes for three representative values of the comoving frequency $\omega$. All quantities in units of the Hawking temperature $T_H=4.6\times10^{-4}$ PHz (in Kelvin $T_H=3.51$ K). The label $(A|B)$ represents a bipartite cut across the modes $A$ and modes $B$ for which entanglement is evaluated.}
\label{table:threshold}
\end{table}

We study how ambient thermal noise in the lab frame degrades the Hawking-generated entanglement and, in particular, determine the threshold temperature above which entanglement is extinguished. This is a rich problem because the threshold temperature depends on the comoving frequency $\omega$. This dependence is easy to understand because the intensity of the Hawking squeezers, $r_H={\rm arctanh} \, e^{-\omega/(2T_H)}$, falls exponentially with $\omega$; hence, entanglement is weaker for modes with large $\omega$ and is consequently more fragile. Moreover, entanglement and its sensitivity to noise also depends on which bipartitions and subsystems of the outgoing state  we consider. We report results for three representative values of the comoving frequency: $\omega/T_H=10^{-1},\, 1, \, 10$. For completeness, we show results obtained from both numerical resolution of the  equations of the model for the dielectric medium \cite{linder16}, as well as from the analytical expressions derived from Eq.~\eqref{eq:SWB}. As shown in Fig.~\ref{fig:ln}, both approaches agree well when computing the entanglement.

In Fig.~\ref{fig:ln}, we plot the LN versus the lab temperature $T_{\rm env}^{\rm (lab)}$ at $\omega/T_H\approx1$ for various subsystems of modes, where, e.g., we use the notation $(k_1|k_4)$ to represent a bipartite cut across the reduced subsystem containing only the Hawking pairs emitted by the white hole horizon; similarly, $(k_1,k_2,k_4|k_3)$ represents the cut partitioning the white hole exterior and black hole exterior, etc. The bipartition with the largest amount of entanglement is $(k_2,k_3,k_4|k_1)$, which partitions the positive norm modes $(k_2,k_3,k_4)$ from the negative norm mode $k_1$. The subystem with the least amount of entanglement is $(k_1|k_2)$, which we do not show on the plot as the entanglement in this subsystem is five orders of magnitude smaller than the rest.

For each subsystem, there exists a threshold lab temperature $T_{\rm env}^{\star\,\rm (lab)}$ at which point entanglement vanishes within that subsystem. The values of $T_{\rm env}^{\star\,\rm (lab)}$ are indicated on Fig.~\ref{fig:ln} by dashed vertical lines. We also give explicit values for $T_{\rm env}^{\star\,\rm (lab)}$ (in units of $T_H=3.51$ K) in Table~\ref{table:threshold} for three representative comoving frequencies $\omega/T_H=10^{-1},1,10$. The threshold temperatures drastically vary across the different subsystems and for different frequencies, which highlights the richness of the entanglement spectrum. Generally though, entanglement is more robust for low frequencies. The lowest  threshold temperature is found for the subsystem $(k_1|k_2)$; the reason being that the entanglement in this subsystem is the weakest, as such entanglement does not originate from the Hawking effect. For various subsystems whose entanglement originates from the Hawking process, and at various frequencies (low and high), we observe threshold temperatures $T_{\rm env}^{\star\,\rm (lab)}/T_H\sim\order{10^3}$ or larger.

\section{Comparison with gravitational black holes}
\label{grav}

\begin{figure}
    \centering
    \includegraphics[width=.5\linewidth]{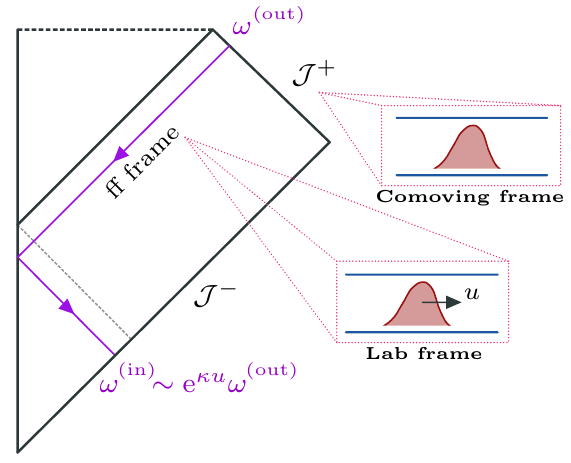}
    \caption{Illustration depicting the correspondence between reference frames for an astrophysical black hole (ff frame and $\mathcal{J}^+$) and an optical analog (lab frame and comoving frame). Blue-shift between ingoing and outgoing frequencies is also shown. For the ff frame, the black hole event horizon is in motion; similarly, for the optical analog,  the white-black hole (i.e., the pulse) is in motion with respect to the lab frame. For observers at $\mathcal{J}^+$, the black hole is stationary; similarly, for the optical analog,  the white-black hole is stationary  in the frame comoving with the pulse.}
    \label{fig:ff_frame}
\end{figure}

The robustness to thermal fluctuations is reminiscent of what happens for astrophysical black holes. In the astrophysical case, the progenitors of the Hawking radiation are modes supported on ultra-high frequencies as measured by observers in the distant past (at past null infinity, $\mathcal{J}_-$). Therefore, any thermal fluctuations in the past would be incapable of significantly affecting the outgoing Hawking spectrum (measured by observers at future null infinity, $\mathcal{J}_+$) or its entanglement with the black hole, unless those thermal fluctuations had an ultra-high temperature (larger than the Planck temperature) in the natural frame at $\mathcal{J}_-$. To better elucidate these points, in both the astrophysical and analog settings, we examine the scenario from the viewpoint of preferred reference frames.

For astrophysical black holes, there are two preferred frames---the ingoing and outgoing frames. The ingoing frame is the natural frame to set up the initial state; it is equivalent to a freely falling (ff) frame (see, e.g., Refs.~\cite{Unruh76,haag1990HawkingRadiation, agullo2010prl}) defined by an observer which crosses the horizon at the same instance the surface of the collapsing object does. The outgoing frame at future null infinity, $\mathcal{J}_+$ is the natural one to evaluate the Hawking radiation. There is an exponential blue-shift between the ingoing and outgoing frequencies (defined in the ingoing and outgoing frames respectively), $\omega^{\rm( in)}\sim e^{\kappa u}\, \omega^{\rm(out)}$, where $\kappa$ is the surface gravity of the black hole (proportional to $T_H$), and $u$ is the retarded time along $\mathcal{J}_+$; see Fig.~\ref{fig:ff_frame} for an illustration.\footnote{Recall that Hawking quanta reach $\mathcal{J}_+$ for asymptotically large $u$.} The blue-shift between these frames has a gravitational origin; wave packets reaching $\mathcal{J}_+$ at late $u$, when propagated back in time towards $\mathcal{J}_-$, come arbitrarily close to the horizon, blue-shifting as they approach the horizon. The trip from the horizon to $\mathcal{J}_-$ does not undo the blue-shift. Thus, it is the time asymmetry induced by gravitational collapse that causes an exponentially large blue-shift. 

The analogs to the ingoing and outgoing frames in the optical systems discussed in this article are the lab and comoving frames, respectively; see Fig.~\ref{fig:ff_frame} for a visual. For the optical system, there is a large blue-shift between the lab frame and comoving, similar to the astrophysical black hole. The difference from the astrophysical case is, of course, the physical origin of the blue-shift. For optical systems, the blue-shift originates from the combined effects of a Lorentz boost and dispersion. Moreover, contrary to the astrophysical case, the blue-shift between lab and comoving frames is not exceedingly (exponentially) large. However, for all practical purposes, it is still large enough to protect the Hawking effect from spurious thermal fluctuations (see Tables~\ref{table:omegas} and~\ref{table:threshold}).

\section{Conclusions}
\label{conclusions}

Optical systems present an intriguing setting to recreate the physics of the Hawking effect. They have attractive advantages, such as current experimental capabilities to manipulate individual photons and to measure their entanglement. These systems also have the peculiar feature that the analog horizons move with respect to the lab frame at a fraction of the speed of light. We have shown in this paper that moving horizons actually provides an advantage: it naturally protects entanglement generated during the Hawking process against background thermal noise. 

The robustness of entanglement to harmful thermal fluctuations derives from an interplay between dispersive effects and the Lorentz boost between the lab frame and the frame comoving with the horizons. Modes with comoving frequency $\omega$ that are of the order of the Hawking temperature $T_H$ appear to be high frequency modes as seen in the lab frame due to the boost. Consequently, the lab temperatures needed to appreciably affect Hawking-generated entanglement are several orders of magnitude larger than $T_H$. 

Such robustness is absent for systems with analog horizons that are at rest in the laboratory frame, like BECs. For BEC analogs, both temperatures $T_H$ and $T_{\rm env}$ are defined in the same frame and, consequently, $T_{\rm env}\approx T_H$ suffices to decohere the outgoing radiation and eliminate all quantum traces of the Hawking effect~\cite{bruschi2013}; in which case, the final state can then be accounted for by a classical  process of amplification of thermal radiation. One is required to work at low temperatures $T_{\rm env}\lesssim T_H$ for quantum features to survive. 

In this sense, optical systems share some analogy with astrophysical black holes. For the latter, the entanglement produced during Hawking's evaporation is shielded against background thermal fluctuations (such as the cosmic microwave background radiation) for which $T_{\rm env}\gg T_H$ for black holes of astrophysical origin. One needs an exponentially high background temperature to noticeably populate the Hawking progenitors and thus decohere astrophysical Hawking particles from their partners. The difference is that, for astrophycial black holes, the protection originates entirely from gravitational blue-shifting; while for optical analogs, it originates from an interplay of blue-shifting (due to a Lorentz boost) and dispersive effects. 

We conclude that entanglement generated by optical analog event horizons is extremely robust to background thermal fluctuations in the lab frame. This observation---together with the fact that precise manipulation of quantum states of light is a staple of modern quantum technologies~\cite{raymer2009,bachor2019guide}---indicates that optical analogs~\cite{philbin08,drori19,rosenberg2020optical,Aguero-Santacruz:2020krw} are promising candidates for observing the genuinely quantum nature of the Hawking effect in a practical setting, hopefully in the near future.

\acknowledgments

We have benefited from  discussions with A. Bhardwaj, D.~Bermudez, M.~Jacquet, D.~Seehy and J.~Wilson. I.A.\ and D.K.\ are supported by the NSF grant PHY-2110273, and by the Hearne Institute for Theoretical Physics. A.J.B. acknowledges discussion with Quntao Zhuang and support from the Defense Advanced Research Projects Agency (DARPA) under Young Faculty Award (YFA) Grant No. N660012014029.

\appendix

\section{Symplectic matrix for white-black hole}\label{app:swb}
To construct the symplectic matrix for the white-black hole, $\bm S_{\rm WB}$ of Eq.~\eqref{eq:SWB}, we can regard the entire scattering process as a succession of four distinct events. [For justification of this, see Refs.~\cite{agullo2022prl,brady2022prd}.]

High-frequency progenitor modes $k_1^{\rm (in)}$ and $k_4^{\rm (in)}$ approach the black hole side of the strong pulse (affected negligibly by a dispersion-induced potential barrier, due to their high frequencies) and participate in a two mode squeezing process converting into Hawking pairs $k_3^{(\rm out)}$ and $k_1^{(\rm int)}$, where $k_3^{(\rm out)}$ denotes the outgoing Hawking radiation and $k_1^{(\rm int)}$ denotes the entangled Hawking partner that falls into the white-black hole. This pair creation event is described by the symplectic matrix
\begin{equation}
\bm S_1= \begin{pmatrix}
\cosh{ r_H}\, \bm{I}_2& 0 & 0 & \sinh{ r_H}\, \bm\sigma_z \\
0 & \bm{I}_2 & 0 & 0 \\
0 & 0 & \bm{I}_2 & 0 \\
\sinh{ r_H}\, \bm\sigma_z & 0 & 0 & \cosh{ r_H}\, \bm{I}_2
\end{pmatrix},
\end{equation}
where the ordering of elements follows the convention in Eq.~\eqref{eq:R_in}. 

As the outgoing Hawking radiation, $k_3^{(\rm out)}$ (a lower frequency mode), propagates outward, it meets a potential barrier where it mixes with the low-frequency mode $k_2^{\rm (in)}$, partly getting scattered back into the white-black hole as $k_2^{\rm (int)}$. This scattering process is described by a symplectic matrix
\begin{align}
    \bm S_2 &= \begin{pmatrix} \bm{I}_2 & 0 & 0 & 0 \\ 
    0 & \cos{\theta_\omega}\bm{I}_2 & 0 &-\sin\theta_\omega\bm{I}_2 \\
    0 & 0 & \bm{I}_2 & 0 \\ 
    0 & \sin\theta_\omega\bm{I}_2 & 0 & \cos{\theta_\omega}\bm{I}_2 \end{pmatrix}, 
    \end{align}
where $\cos^2\theta$ is the greybody factor of the potential barrier (i.e., its transmission probability). Taken together, $\bm S_2\bm S_1$ denotes the dynamics induced by the black hole event horizon.

Following the scattering processes induced by black hole, we have the white hole dynamics. Intuitively, the white hole dynamics follows the black hole dynamics because the black hole pair creation mechanism actually sources the white hole scattering processes. Indeed, the back-scattered mode $k_2^{\rm (int)}$ traverses through the interior of the white-black hole unscathed. However, as it exits the white hole event horizon, it meets the $k_3^{\rm(in)}$ mode (the time-reverse of the outgoing Hawking radiation) at a potential barrier and scatters. This scattering process is described by the matrix
     \begin{equation}  \bm S_3 = \begin{pmatrix} \bm{I}_2 & 0 & 0 & 0 \\
    0 & \cos{\theta_\omega}\bm{I}_2 & \sin\theta_\omega\bm{I}_2 & 0 \\
    0 & -\sin\theta_\omega\bm{I}_2 & \cos{\theta_\omega}\bm{I}_2 & 0 \\
    0 & 0 & 0 & \bm{I}_2 \end{pmatrix},
\end{equation}
Observe the sign reversal of the angle $\theta$, which is due to the fact that, for symmetric pulses, the white hole can be understood as the time-reverse of the black hole.

The portion of ingoing mode $k_3^{\rm (in)}$ that makes it to the white-hole event horizon then interacts with the $k_1^{\rm (int)}$ from the black hole pair creation process by a two-mode squeezing interaction,\footnote{Locally, the $k_1^{\rm (int)}$ mode occupies a thermal state at temperature $T_H$; thus, the white hole pair creation process is stimulated by the Hawking partners generated by the black hole.}
\begin{equation}
    \bm S_4 = \begin{pmatrix}
    \cosh{ r_H}\bm{I}_2 & 0  & -\sinh{ r_H}\bm\sigma_z & 0 \\
    0 & \bm{I}_2 & 0 & 0 \\
    0 & 0  & 0 & \bm{I}_2 \\
    -\sinh{ r_H}\bm\sigma_z & 0 & \cosh{ r_H}\bm{I}_2 & 0 
\end{pmatrix}.
\end{equation}
 Again, the relative sign between  $S_1$ and $S_4$ is due to the inverse character of the white hole relative to the black hole. The product $\bm S_3\bm S_4$ denotes the white hole scattering process.

Altogether, $\bm S_{\rm WB}=\bm S_4\bm S_3\bm S_2\bm S_1$, which reduces to Eq.~\eqref{eq:SWB}; see Fig.~\ref{fig:modestructure} for a visual aid and Fig.~\ref{fig:wbh_circuit} for a circuit description of the processes. 

\section{Mean number of quanta}\label{app:particle_number}

When analyzing particle pair production in analog gravity or quantum fields in curved spacetime, one often computes the mean number of quanta produced. For completeness, we do so here for the optical white-black hole pair and an thermal initial state.   

Quite generally, starting from the output covariance matrix $\bm\sigma^{\rm (out)}$, we can compute the outgoing number of quanta in the mode $k_i^{\rm (out)}$ via 
\begin{equation}
    \ev{\hat{n}^{\rm(out)}_i}=\frac{1}{4}\text{Tr}\{\bm\sigma_i^{\rm(red)}\}+\frac{1}{2}\bm{\mu}^{\rm(red)\, \top}\bm{\mu}^{\rm(red)} -\frac{1}{2}\, .\label{eq:mean_quanta}
\end{equation}    
where $\bm{\mu}_i^{\rm(red)}$ and $\bm\sigma_i^{\rm(red)}$ are the sub-vector and sub-matrix of $\bm{\mu}^{\rm(out)}$ and $\bm{\sigma}^{\rm(out)}$ for the mode $k_i$.\footnote{The relation between the mean number of quanta and the first and second moments [Eq.~\eqref{eq:mean_quanta}] holds for any quantum state, not just Gaussian states. Though the state under consideration is indeed a Gaussian state.} Assuming that the modes are initially populated with thermal quanta and utilizing Eqs.~\eqref{eq:mu_out_wb} and~\eqref{eq:sigma_out_wb}, it is straightforward to obtain the following analytical expressions for the outgoing quanta in each mode. First, for brevity, define $N_{\rm env;k}\equiv(1+2 n_{\rm env;k})$. Then,
\begin{multline}
 \ev{\hat{n}^{\rm(out)}_1} = \frac{1}{2} \bigg[-1+ N_{\rm env;1} \cosh^4r_H+ \sinh^2r_H \Big[\cosh^2 r_H \Big(N_{\rm env;4} \cos^4\theta  - 2  N_{\rm env;1} \sin^2\theta\Big)\, \\+  \cos^2\theta\,  \Big(N_{\rm env;3} +N_{\rm env;2} \sin^2\theta\Big) \Big] + N_{\rm env;1} \sin^4\theta \sinh^4r_H\bigg],
\end{multline}
\begin{multline}
\ev{\hat{n}^{\rm(out)}_2}=\frac{1}{2} \bigg[-1 +N_{\rm env;2} \cos^4\theta+   N_{\rm env;3} \sin^2\theta  +\cos^2\theta \sin^2\theta \Big( N_{\rm env;4} \cosh^2r_H+ N_{\rm env;1} \sinh^2r_H\Big)\bigg],
\end{multline}
\begin{equation}
\ev{\hat{n}^{\rm(out)}_3}=\frac{1}{2} \bigg[-1+ N_{\rm env;2} \sin^2\theta+\cos^2\theta \Big(N_{\rm env;4} \cosh^2r_H+N_{\rm env;1} \sinh^2r_H\Big)\bigg],\quad\quad\quad\quad\,\,
\end{equation}
\begin{multline}
\ev{\hat{n}^{\rm(out)}_4}=\frac{1}{2} \bigg[-1+ N_{\rm env;4} \cosh^4r_H \sin^4\theta+N_{\rm env;4} \sinh^4r_H +\cosh^2r_H \bigg(\cos^2\theta\, \Big[N_{\rm env;3}+N_{\rm env;2}\sin^2\theta\Big]\\ +\sinh^2r_H \Big[N_{\rm env;1} \cos^4\theta-2 N_{\rm env;4} \sin^2\theta\Big]\bigg)\bigg].
\end{multline}
If we substitute $n_{\rm env;i}=0$ ($N_{\rm env;i}=1$) into this expression, such that the initial state is vacuum, we obtain the radiation spontaneously emitted by the system. Terms proportional to $n_{\rm env;i}$ thus correspond to induced or stimulated radiation.




%

\end{document}